\documentclass[a4paper]{revtex4}
\usepackage{setspace}
\usepackage{amssymb}
\usepackage{amsfonts}

\usepackage{graphicx}
\usepackage{dcolumn}
\usepackage{bm}
\begin{document}

\markboth{S. D. Campos}
{Phenomenological Analysis of $pp$ and $\bar{p}p$ Elastic Scattering}

%
%

\title{PHENOMENOLOGICAL ANALYSIS OF $PP$ AND $\bar{P}P$ ELASTIC SCATTERING DATA IN THE IMPACT PARAMETER SPACE}

\author{S. D. CAMPOS\footnote{Permanent Address.}} \email{sergiodc@ufscar.br.br}
\affiliation{Universidade Federal de S\~ao Carlos, campus de Sorocaba\\
18052780, Sorocaba, S\~ao Paulo, Brazil}


\date{\today}


\begin{abstract}
We use an almost model-independent analytical parameterization for $pp$ and $\bar{p}p$ elastic scattering data to analyze the eikonal, profile, and inelastic overlap functions in the impact parameter space. Error propagation in the fit parameters allows estimations of uncertainty regions, improving the geometrical description of the hadron-hadron interaction. Several predictions are shown and, in particular, the prediction for $pp$ inelastic overlap function at $\sqrt{s}=14$ TeV shows the saturation of the Froissart-Martin bound at LHC energies.

\end{abstract}


\maketitle



\section{Introduction}\label{intro}
The general belief that Quantum Chromodynamics, QCD, is $the$ correct theory of strong interactions being the only one able to describe this physical regime. However, reliable predictions can actually be extracted for a very restricted class of physical quantities, i.e., in general QCD can only handle hard processes (perturbative processes). Soft processes (high-energy elastic hadron scattering, for example) are usually described by models and some of them are "QCD inspired" \cite{qcdinspired}. However, there are models reaching remarkable results as, for example, Cheng and Wu model \cite{chengwu} (pre-QCD) that asymptotically predicts the total cross section, $\sigma_{tot}(s)$, will increase as $R(s)$ where $R(s)$ is a typical interaction radius which grows logarithmically with energy, $\sigma_{tot}(s)\sim \log^2{s}$. 

Actually, explain high-energies $pp$ and $\bar{p}p$ elastic scattering is a very difficult task since it is necessary to understand the formally theoretical results obtained within Axiomatic Quantum Field Theory, as well as the experimental available data from hadron-hadron collision experiments realized over decades. In this way, models of elastic scattering must obey theorems and bounds as the Froissart-Martin bound, Pumplin bound, Pomeranchuk theorem, and others since these theoretical results are model-independent. On the other hand, these formal results do not allow us to write a functional form for $pp$ and $\bar{p}p$ elastic scattering amplitude. Consequently, the experimental data play a central role in this kind of analysis. Nowadays, experimental data about $pp$ and $\bar{p}p$ elastic scattering allows a novel understanding of scattering reactions because its data set are the most complete and that reaches the highest energies at colliders. The total cross section is an example of the importance of experimental data analysis since before the advent of ISR collider, it was assumed that the $pp$ and $\bar{p}p$ cross sections were approaching a common constant value of about 40 mb, at $\sqrt{s}\rightarrow\infty$. When the ISR was turned on, some results showed that $pp$ cross section was rising with energy, more precisely, had a rise that could be fitted with a $\log^2{s}$ leading term. 

In \cite{sdc} a phenomenological approach to explain high energies $pp$ and $\bar{p}p$ total cross sections, differential cross sections and $\rho$ parameter is presented. This parameterization is based on bounds and theorems of the Axiomatic Quantum Field Theory and on $pp$ and $\bar{p}p$ experimental available data above $\sqrt{s}=20$ GeV for the total cross sections, differential cross sections, and $\rho$ parameter. This almost model-independent approach may be improved with additional theoretical information as well as using future experimental measurements at RHIC, LHC, and TEVATRON colliders allowing more precise descriptions. Furthermore, predictions in the momentum transfer space were performed and compared with those in the literature. In \cite{sdc} a more detailed discussion is presented.

On the other hand, there exist a classical description of the scattering amplitude based on the impact parameter representation, which the corresponding amplitude is perpendicular to the beam direction and thus is the same in the laboratory and center-of-mass system. The elastic scattering amplitude in the momentum transfer space allows, in the impact parameter space, an estimate of the interacting particle sizes and an analysis of their internal structures. There are several papers about this theoretical approach in the literature with pros and cons \cite{seminal}. In particular, Amaldi and Schubert had performed a comprehensive and coherent analysis of $pp$ data at ISR energies, which had shown that the rise of the total cross section is a peripheral phenomenon \cite{amaldi}.  

Using the well established parameterization taken from \cite{sdc} we investigate here the eikonal, profile, and inelastic overlap functions using the impact parameter representation via the Fourier-Bessel transform of the elastic scattering amplitude. Numerical values of the fit parameters used here were obtained in the momentum transfer space from simultaneous fits for the total, differential cross section and $\rho$ parameter and then used to describe eikonal, profile and overlap functions in the impact parameter space. Standard error propagation in the fit parameters is also performed allowing more realistic descriptions of these functions. 

This paper is organized as follows. In section \ref{basic}, the basic formalism of the scattering amplitude in the impact parameter representation is presented. In section \ref{results} we show the numerical values of the parameters used here and present the results for the eikonal, profile and inelastic overlap functions. In Section \ref{predictions} predictions to other energies not used in the fitting procedures are shown, and in section \ref{final} we present our final remarks.  

\section{Basic Formalism}\label{basic}
We consider elastic scattering amplitude in a simple geometrical picture using an impact parameter $b$ and the Fourier-Bessel transform of the transferred momentum $-t=q^2$. In this picture, the profile function is introduced in high energy diffractive processes to describe the shape of the collision partners in the plane transverse to the beam axis. Assuming that longitudinal momentum transfer is negligible the profile function can be written in the impact parameter representation as \cite{renk,predazzi}

\begin{eqnarray}
\label{espa2}F(s,q^2)=i{\int_0}^{\infty}bdbJ_0(qb)[1-e^{i{\chi}(s,b)}]=i{\int_0}^{\infty}bdbJ_0(qb)\mathrm{\Gamma}(s,b),
\end{eqnarray}

\noindent where $\sqrt{s}$ is the energy in the center-of-mass system, $J_0$ is a zeroth order Bessel function, and $\chi(s,b)$ is the eikonal function written as

\begin{eqnarray}
\label{eikonal}\chi(s,b)=\mathrm{Re}\chi(s,b)+i\mathrm{Im}\chi(s,b).
\end{eqnarray}

The eikonal correspond to the continuum (complex) phase shift and in both the high energy limit and the semi-classical approximation, $\chi(s,b)=2\delta(s,b)$. Moreover, $\mathrm{Im}\chi(s,b)$ is the so-called opacity function related to the matter distribution inside the incident particles. Explicitly, the real and imaginary parts of the eikonal function can be expressed in terms of the real and imaginary parts of the profile function $\mathrm{\Gamma}(s,b)$ as 

\begin{eqnarray}
\mathrm{Re}\chi(s,b)=\tan^{-1}\left[\frac{\mathrm{Im}\mathrm{\Gamma}(s,b)}{\mathrm{Re}\mathrm{\Gamma}(s,b)-1}\right], 
\end{eqnarray}

\begin{eqnarray}
\mathrm{Im}\chi(s,b)=\ln\left[\frac{1}{\sqrt{\mathrm{Im}^2\mathrm{\Gamma}(s,b)+[1-\mathrm{Re}\mathrm{\Gamma}(s,b)]^2}}\right].
\end{eqnarray} 

The unitarity condition in impact parameter representation sets a link among total ($\sigma_{tot}$), elastic ($\sigma_{el}$) and inelastic ($\sigma_{inel}$) differential cross sections. Consequently, it gives a remarkable opportunity to study at the same time both elastic and inelastic properties of the interaction. These physical quantities can be written in terms of $\mathrm{\Gamma}(s,b)$ as \cite{predazzi}  

\begin{eqnarray}
\label{eqmarcio1} \sigma_{el}(s)=\int d^2b|\mathrm{\Gamma}(s,b)|^2, 
\end{eqnarray}

\begin{eqnarray}
\label{eqmarcio2} \sigma_{inel}(s)=\int d^2b[2\mathrm{Re}\mathrm{\Gamma}(s,b)-|\mathrm{\Gamma}(s,b)|^2], 
\end{eqnarray}

\begin{eqnarray}
\label{eqmarcio3} \sigma_{tot}(s)=2\int d^2b\mathrm{Re}\mathrm{\Gamma}(s,b).
\end{eqnarray}

In the impact parameter representation unitarity condition can be written as

\begin{eqnarray}
\label{unitaridade} 2\mathrm{Re}\mathrm{\Gamma}(s,b)=|\mathrm{\Gamma}(s,b)|^2+G_{inel}(s,b),
\end{eqnarray}

\noindent where $G_{inel}(s,b)$ is the (real) inelastic overlap function. We can relate both overlap and inelastic cross section by

\begin{eqnarray}
\label{ginelmarcio} \sigma_{inel}(s)=\int_0^\infty d^2b G_{inel}(s,b). 
\end{eqnarray} 

In (\ref{unitaridade}), $|\mathrm{\Gamma}(s,b)|^2$ represents the shadow contribution of elastic channel, and it connects both elastic and inelastic channels. Using the constraint $\mathrm{Im}\chi(s,b)\geq 0$, it is easy to see that $G_{inel}(s,b)$ represents the probability of an elastic scattering in $(s,b)$ given by 

\begin{eqnarray}
\label{eq10} G_{inel}(s,b)=1-e^{-2\mathrm{Im}\chi(s,b)}\leq{1}.
\end{eqnarray}

In general, derivation of the above formulas follows the usual argument which tends to imply that the impact parameter is a high energy, small angle approximation. However, using the Watson-Sommerfeld transform, Islam proves that the impact parameter representation is an exact representation of the scattering amplitude, valid for all energies and angles \cite{islam}.  

The basic assumptions about the almost model-independent analytical parameterization for $pp$ and $\bar{p}p$ elastic scattering amplitude was well established in \cite{sdc}. Besides those assumptions, we assume that $F(s,q^2)$ is also valid to $q^2\rightarrow{\infty}$ obtaining a unique $\mathrm{\Gamma(s,b)}$ (Appendix A). 

In the momentum transfer space, the elastic scattering amplitude for $pp$ and $\bar{p}p$ can be written as

\begin{eqnarray}
\label{espalhamentopp}\frac{F_{pp}(s,q^2)}{s}=\frac{K}{s}+\sum_{i=1}^n \left\{
\frac{\pi}{2}\left[\alpha_i'(s)-\alpha_i(s)\beta_i'(s)q^2\right]
e^{-\beta_i(s) q^2} \right. \nonumber \\
+\left.\frac{\pi}{4}
\left[\alpha_i(s) e^{-\beta_i(s)
    q^2}-\bar{\alpha_i}(s)e^{-\bar{\beta_i}(s) q^2}\right]\right\}+i\sum_{i=1}^n\alpha(s)e^{-\beta(s)q^2},
\end{eqnarray}

\begin{eqnarray}
\label{espalhamentopbp}\frac{F_{\bar{p}p}(s,q^2)}{s}=\frac{K}{s}+\sum_{i=1}^n\left\{
\frac{\pi}{2}\left[\bar{\alpha_i}'(s)-\bar{\alpha_i}(s)\bar{\beta_i}'(s)q^2\right]e^{-\bar{\beta_i}(s) q^2} \right. \nonumber \\
-\left.\frac{\pi}{4}
\left[\alpha_i(s) e^{-\beta_i(s)
    q^2}-\bar{\alpha_i}(s)e^{-\bar{\beta_i}(s) q^2}\right]\right\}+i\sum_{i=1}^n\bar{\alpha}(s)e^{-\bar{\beta}(s)q^2}.
\end{eqnarray} 

\noindent where $K$ is the subtraction constant, $\alpha_i(s)=A_i+B_i\ln s+C_i\ln^2 s$, $\bar{\alpha}_i(s)=\bar{A}_i+\bar{B}_i\ln s+\bar{C}_i\ln^2 s$, $\beta_i(s)=D_i+E_i\ln s$, $\bar{\beta}_i(s)=\bar{D}_i+\bar{E}_i\ln s$, and $A_i$, $\bar{A}_i$, $B_i$, $\bar{B}_i$, $C_i$, $\bar{C}_i$, $D_i$, $\bar{D}_i$, $E_i$, and $\bar{E}_i$ are fit parameters. 

Using the Fourier-Bessel transform, the above elastic scattering amplitude provides the real and imaginary parts of the $pp$ and $\bar{p}p$ profile function 

\begin{eqnarray}
\label{p60}\mathrm{Re}\mathrm{\Gamma}(s,b)_{pp}=\frac{1}{4\pi s}\sum_{i=1}^3{\frac{\alpha_i(s)}{2\beta_i(s)}}e^{\frac{-b^2}{4\beta_i(s)}},
\end{eqnarray}

\begin{eqnarray}
\nonumber\mathrm{Im}\mathrm{\Gamma}(s,b)_{pp}=-\frac{1}{8}\sum_{i=1}^3\left[{\frac{\alpha'_i(s)}{2\beta_i(s)}-\frac{\alpha_i(s)\beta'_i(s)}{8\beta_i^3(s)}(4\beta_i(s)-b^2)}\right]e^{\frac{-b^2}{4\beta_i(s)}}-\\
\label{p70}-\frac{1}{32}\sum_{i=1}^3\left[\frac{\alpha_i(s)}{2\beta_i(s)}e^{\frac{-b^2}{4\beta_i(s)}}-\frac{\bar{\alpha}_i(s)}{2\bar{\beta}_i(s)}e^{\frac{-b^2}{4\bar{\beta}_i(s)}}\right].
\end{eqnarray}

\begin{eqnarray}
\label{p100}\mathrm{Re}\mathrm{\Gamma}(s,b)_{\bar{p}p}=\frac{1}{4\pi s}\sum_{i=1}^3{\frac{\bar{\alpha}_i(s)}{2\bar{\beta}_i(s)}}e^{\frac{-b^2}{4\bar{\beta}_i(s)}},
\end{eqnarray}

\begin{eqnarray}
\nonumber\mathrm{Im}\mathrm{\Gamma}(s,b)_{\bar{p}p}=-\frac{1}{8}\sum_{i=1}^3\left[{\frac{\bar{\alpha}'_i(s)}{2\bar{\beta}_i(s)}-\frac{\bar{\alpha}_i(s)\bar{\beta}'_i(s)}{8\bar{\beta}_i^3(s)}(4bar{\beta}_i(s)-b^2)}\right]e^{\frac{-b^2}{4\bar{\beta}_i(s)}}+\\
\label{p80}+\frac{1}{32}\sum_{i=1}^3\left[\frac{\alpha_i(s)}{2\beta_i(s)}e^{\frac{-b^2}{4\beta_i(s)}}-\frac{\bar{\alpha}_i(s)}{2\bar{\beta}_i(s)}e^{\frac{-b^2}{4\bar{\beta}_i(s)}}\right].
\end{eqnarray}

Notice that the real (imaginary) part of the profile function is connected with the imaginary (real) part of the elastic scattering amplitude. Therefore, if at high-energies, $s\rightarrow \infty$, the elastic scattering amplitude (momentum transfer space) is imaginary dominant, then the profile function (impact parameter space) will be real dominant.

In the impact parameter representation, the blackness growth speed is about zero at $b=0$, leading saturation of the unitarity limit. Therefore, if imaginary part of the profile function can be neglected, then imaginary part of the scattering amplitude tends asymptotically to a limit, i.e., it cannot grow arbitrarily fast with the energy increase. As well established, the optical theorem in the momentum transfer space connects the imaginary part of the forward elastic scattering and total cross section, hence it is possible at future LHC energies obtain clear indications of saturation of the Froissart-Martin bound.

\section{Fit Results}\label{results}
The efficient description of the global elastic scattering data with an economical number of free parameters is a very difficult task since almost relevant physical quantities depend on both real and imaginary parts of $F(s,q^2)$. Furthermore, model free parameters are, in general, so intricate that practically avoid identifying the explicit role of each parameter in the description of the experimental data and, in most of the cases, numerical calculations may introduce bias or loss of information, and does not allow standard error propagation. Therefore, the use of empirical or model-independent information about the physical quantities that characterize the elastic scattering amplitude is crucial to minimize these problems. 

Keeping in mind the above explanation, the almost-model independent parameterization proposed in \cite{sdc} reaches global and efficient description of the elastic scattering data. This parameterization, although entail many parameters (30 free parameters to be exact), allows a global analysis of all elastic experimental data ($\sigma_{tot}$, $\rho$, and differential cross section), allowing us to determine the uncertainties of the fit parameters and take into account the error propagation with standard methods. Furthermore, simultaneous fits for $pp$ and $\bar{p}p$ experimental data in the forward direction as well as including data beyond the forward direction may result in a predictive formalism in both energy and momentum transfer \cite{sdc}.

The numerical values of the fit parameters obtained in the momentum transfer space are used to describe the profile, eikonal, and inelastic overlap functions in the impact parameter space. It is expected that the predictive character of these parameters remain in the impact parameter representation. In Table \ref{table1} the fit parameters obtained from data up to $q^2_{max}=2.0$ GeV$^2$ and up to $q^2_{max}=14.0$ GeV$^2$ are displayed. For the sake of the simplicity, only profile, eikonal, and inelastic overlap functions achieved from data up to $q^2_{max}=14.0$ GeV$^2$ are shown.

\begin{table*}[h]
\caption{Results to $pp$ and $\bar{p}p$ simultaneous fits to $d\sigma/dq^2$, $\sigma_{tot}$, $\rho$ at $q^2_{max}=2.0$ GeV$^2$ and $q^2_{max}=14.0$ GeV$^2$, where $K = -0.1053 \pm 0.0048$ and $K = 49.7 \pm 1.7$, respectively. All parameters in GeV$^{-2}$, $K = -0.1053 \pm 0.0048$ and $C_1=\bar{C_1}+\bar{C_2}+\bar{C_3}-C_2-C_3$.}
{\begin{tabular}{@{}ccccccc@{}} \toprule
\noalign{\smallskip}
\multicolumn{2}{c}{$pp$}  & & &\multicolumn{2}{c}{$\bar{p}p$}  \\
\noalign{\smallskip}\hline\noalign{\smallskip}
& $q^2_{max}=2.0$ GeV$^2$ & $q^2_{max}=14.0$ GeV$^2$ & & $q^2_{max}=2.0$ GeV$^2$ & $q^2_{max}=14.0$ GeV$^2$\\
\noalign{\smallskip}\hline\noalign{\smallskip}
$A_1$&91.13$\pm$0.28            &109.70$\pm$0.28     &$\bar{A}_1$&119.61$\pm$0.44 &  112.28   $\pm$0.44 \\
$B_1$&-12.939$\pm$0.039         &-16.529$\pm$0.039   &$\bar{B}_1$&-2.486$\pm$0.073& -0.468   $\pm$0.074 \\
$C_1$&                          &                    &$\bar{C}_1$&-0.0174$\pm$0.0038& -0.1673  $\pm$0.0039 \\
$D_1$&-7.79$\pm$0.33            &-8.91$\pm$0.32      &$\bar{D}_1$&3.134$\pm$0.067& 3.170    $\pm$0.069 \\
$E_1$&2.908$\pm$0.051           &3.045$\pm$0.050     &$\bar{E}_1$&0.4884$\pm$0.0078& 0.4860   $\pm$0.0082 \\
$A_2$&16.82$\pm$0.22            &-4.06$\pm$0.23      &$\bar{A}_2$&14.51$\pm$0.15 & 10.23    $\pm$0.15 \\
$B_2$&7.071$\pm$0.030           &11.387$\pm$0.030    &$\bar{B}_2$&-6.730$\pm$0.024& -6.756   $\pm$0.027 \\
$C_2$&0.3027$\pm$0.0047         &0.0952$\pm$0.0047   &$\bar{C}_2$&0.9035$\pm$0.0027& 0.9613   $\pm$0.0029 \\
$D_2$&1.647$\pm$0.014           &1.290$\pm$0.014     &$\bar{D}_2$&-1.549$\pm$0.011& -1.476   $\pm$0.013 \\
$E_2$&0.4646$\pm$0.0022         &0.5097$\pm$0.0023   &$\bar{E}_2$&0.5521$\pm$0.0011& 0.5645   $\pm$0.0012 \\
$A_3$&0.2582$\pm$0.0049         &1.0554$\pm$0.0077   &$\bar{A}_3$&-17.160$\pm$0.055 & -8.148   $\pm$0.031 \\
$B_3$&-0.09894$\pm$0.00074      &-0.3607$\pm$0.0013  &$\bar{B}_3$&2.6083$\pm$0.0060& 1.2313   $\pm$0.0040 \\
$C_3$&0.5921E-02$\pm$0.0070E-02 &0.02372$\pm$0.00012 &$\bar{C}_3$&-0.11298$\pm$0.00038& -0.05572 $\pm$0.00025 \\
$D_3$&0.303$\pm$0.0019          &0.6454$\pm$0.0081   &$\bar{D}_3$&1.5672 $\pm$0.0084 & 0.9272   $\pm$0.0072 \\
$E_3$&0                         &0.0176$\pm$0.0011   &$\bar{E}_3$&0&0.03859  $\pm$0.0011 \\
\noalign{\smallskip}
\hline
\end{tabular}
\label{table1}}
\end{table*}


If we use the Fourier-Bessel transform, then the elastic scattering amplitude in the momentum transfer space implies in analytic functions in the impact parameter space. Therefore, error propagation in the fit parameters allows the estimation of the uncertainty regions \cite{errorprop}, i.e., the uncertainties in the free parameters have been propagated to the profile, eikonal and inelastic overlap functions in the impact parameter space. We may estimate the confidence region associated with all these functions by adding and subtracting the corresponding uncertainties. We choose not to show the error bands as thatched areas since there are sometimes very narrow bands indeed turning this kind of representation useless here.  

\subsection{Profile function}\label{profile}
\begin{figure*}[ht]
\centering
\includegraphics[width=12.0cm, height=10.0cm]{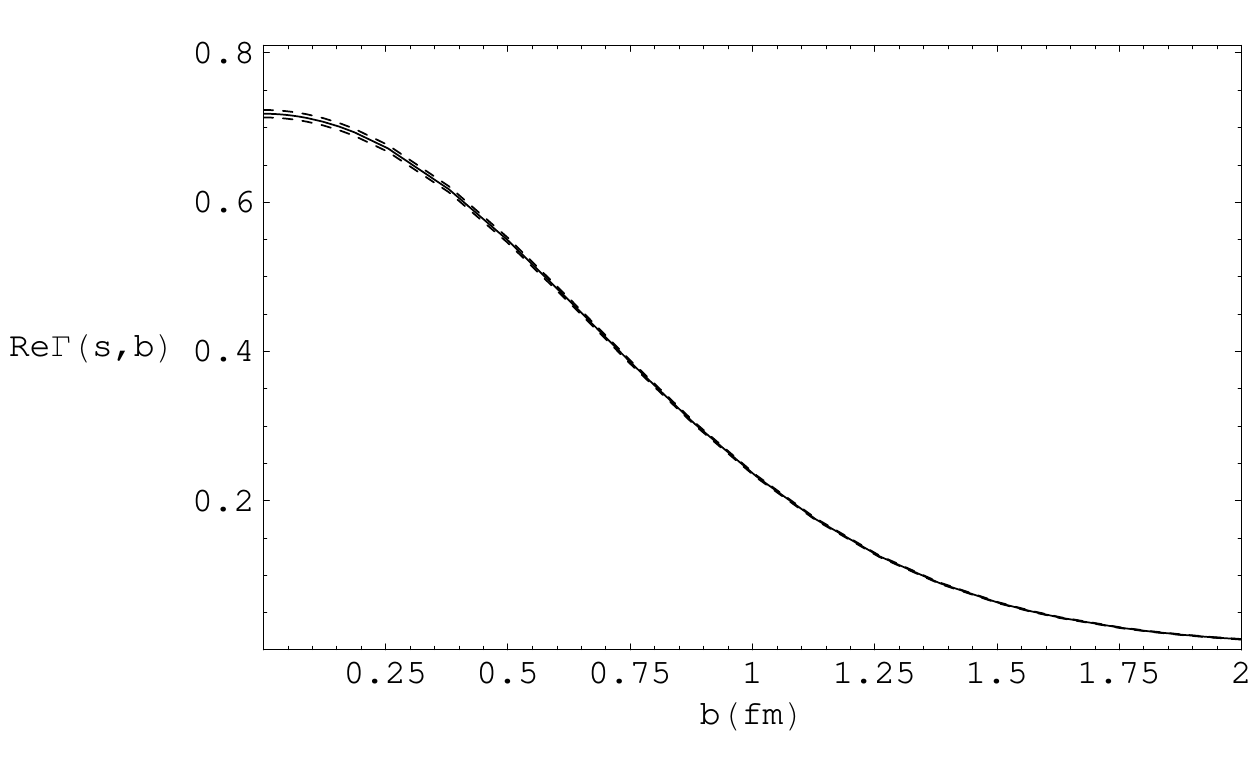}
\includegraphics[width=12.0cm, height=10.0cm]{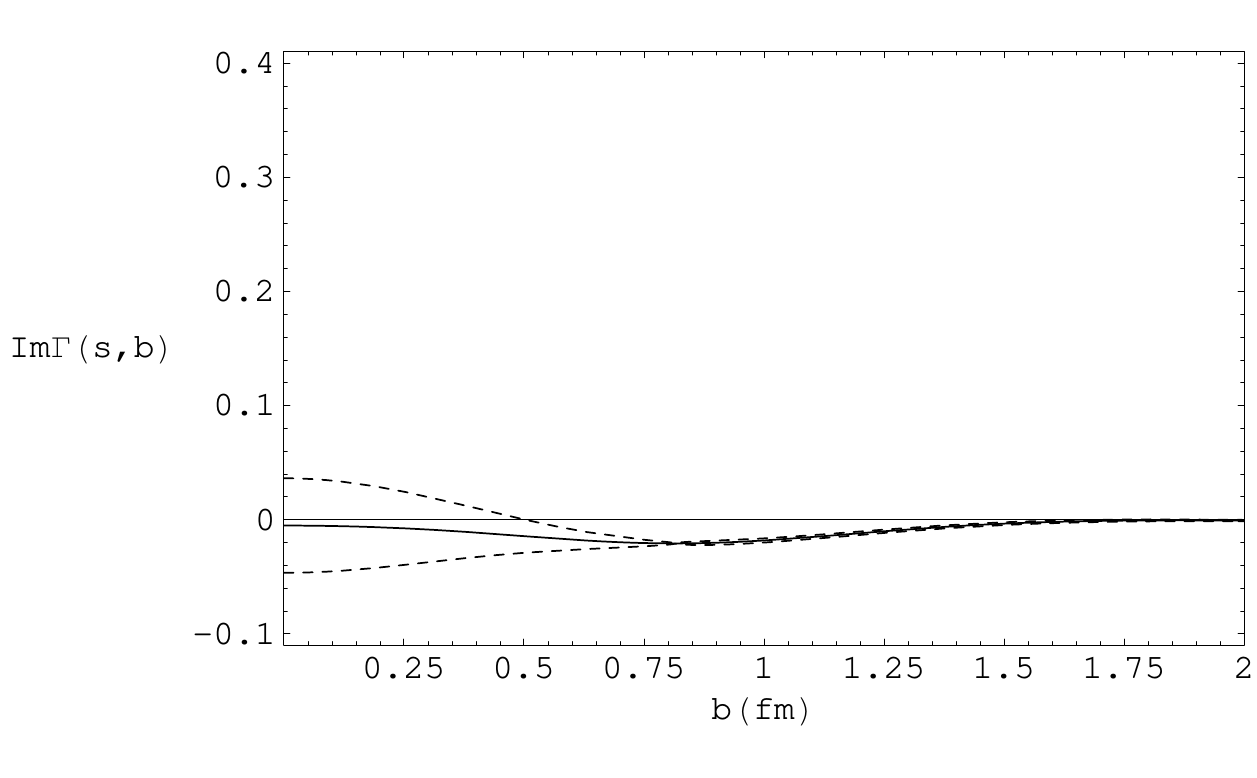}
\vspace*{8pt}
\caption{Real e imaginary parts of the profile function to $pp$ at $\sqrt{s}=52.8$ GeV. Results obtained from the fit parameters to
  ${q^2}_{max}=14.0$ GeV$^2$. Dashed lines represent the error propagation.}
\label{perfil2}
\end{figure*}

\begin{figure*}[ht]
\centering
\includegraphics[width=12.0cm, height=10.0cm]{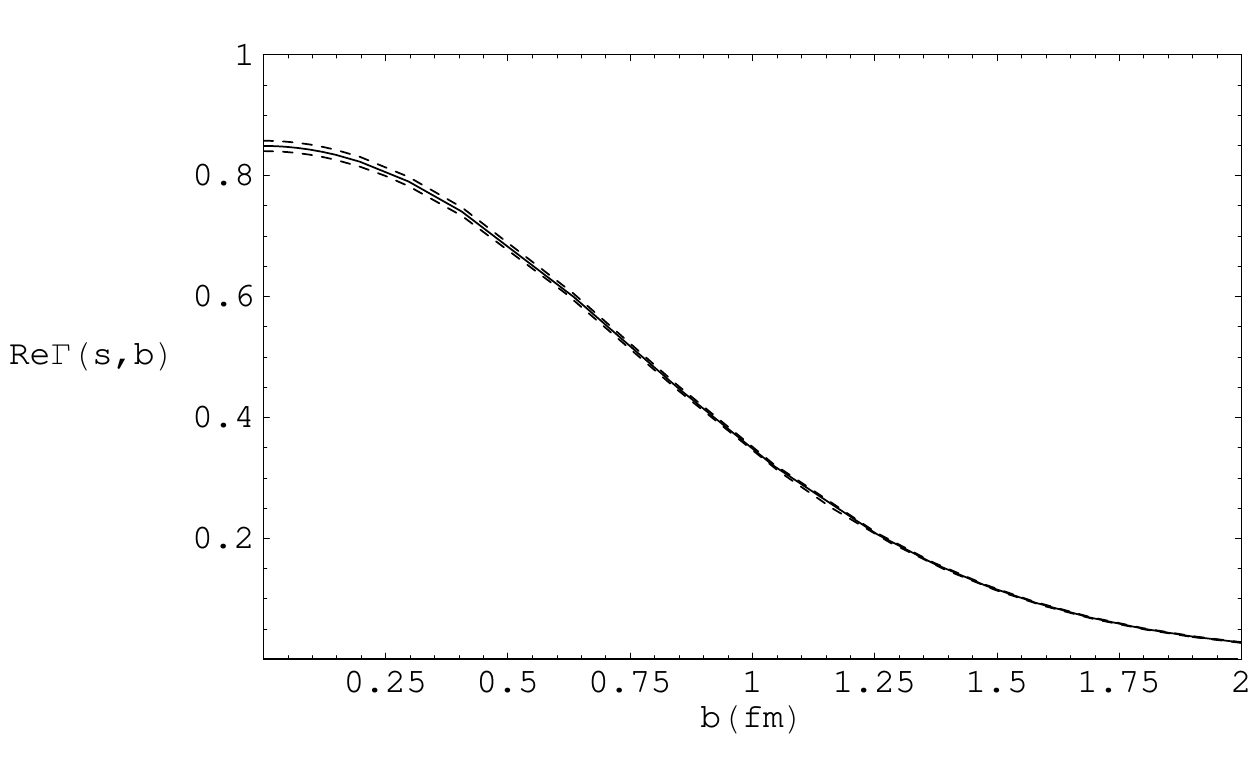}
\includegraphics[width=12.0cm, height=10.0cm]{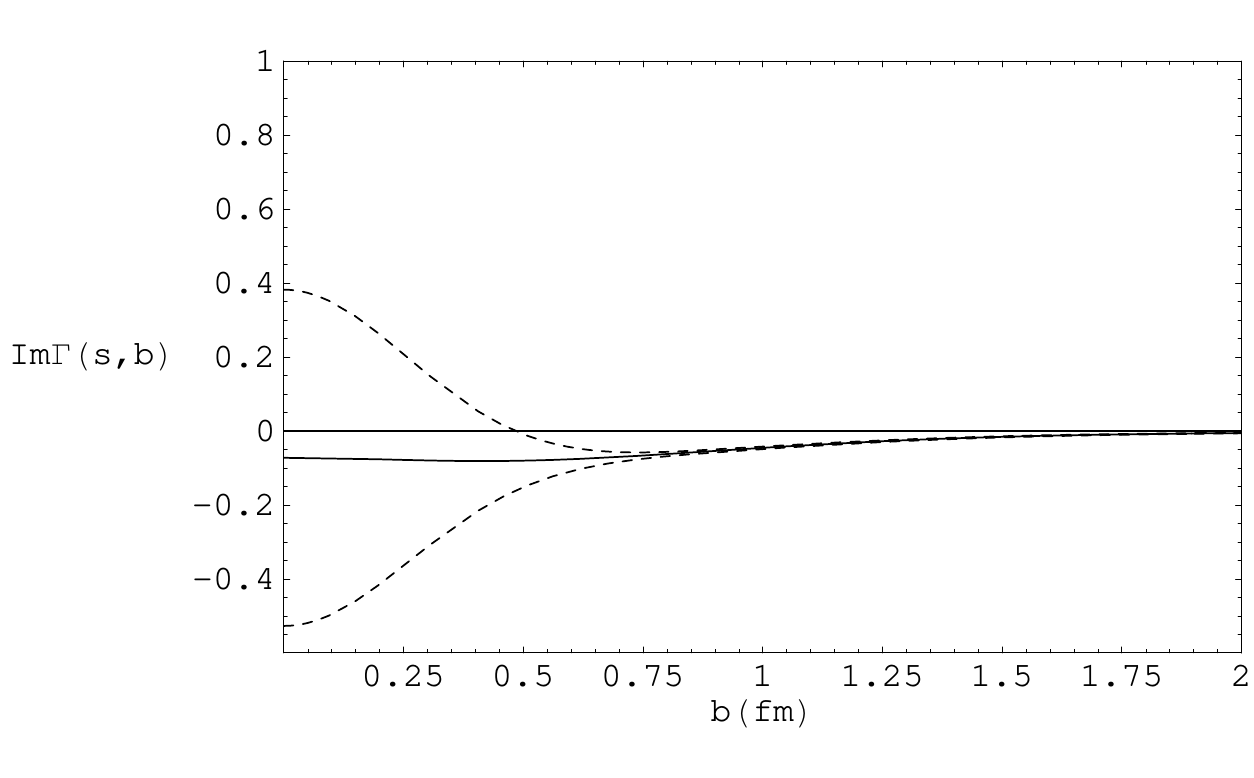}
\vspace*{8pt}
\caption{Real e imaginary parts of the profile function to $\bar{p}p$ at $\sqrt{s}=546.0$ GeV. Results obtained from the fit parameters to
  ${q^2}_{max}=14.0$ GeV$^2$. Dashed lines represent the error propagation.}
\label{perfil11}
\end{figure*}

\begin{figure*}[ht]
\centering
\includegraphics[width=12.0cm, height=10.0cm]{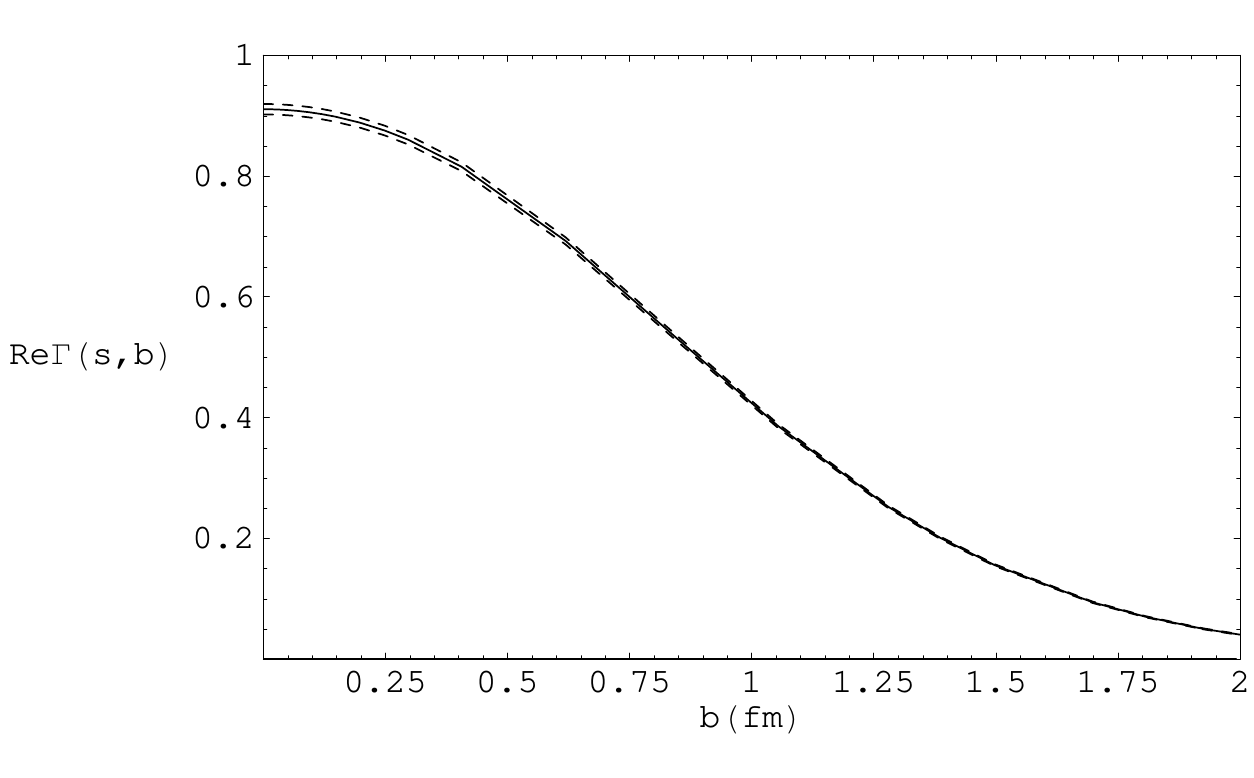}
\includegraphics[width=12.0cm, height=10.0cm]{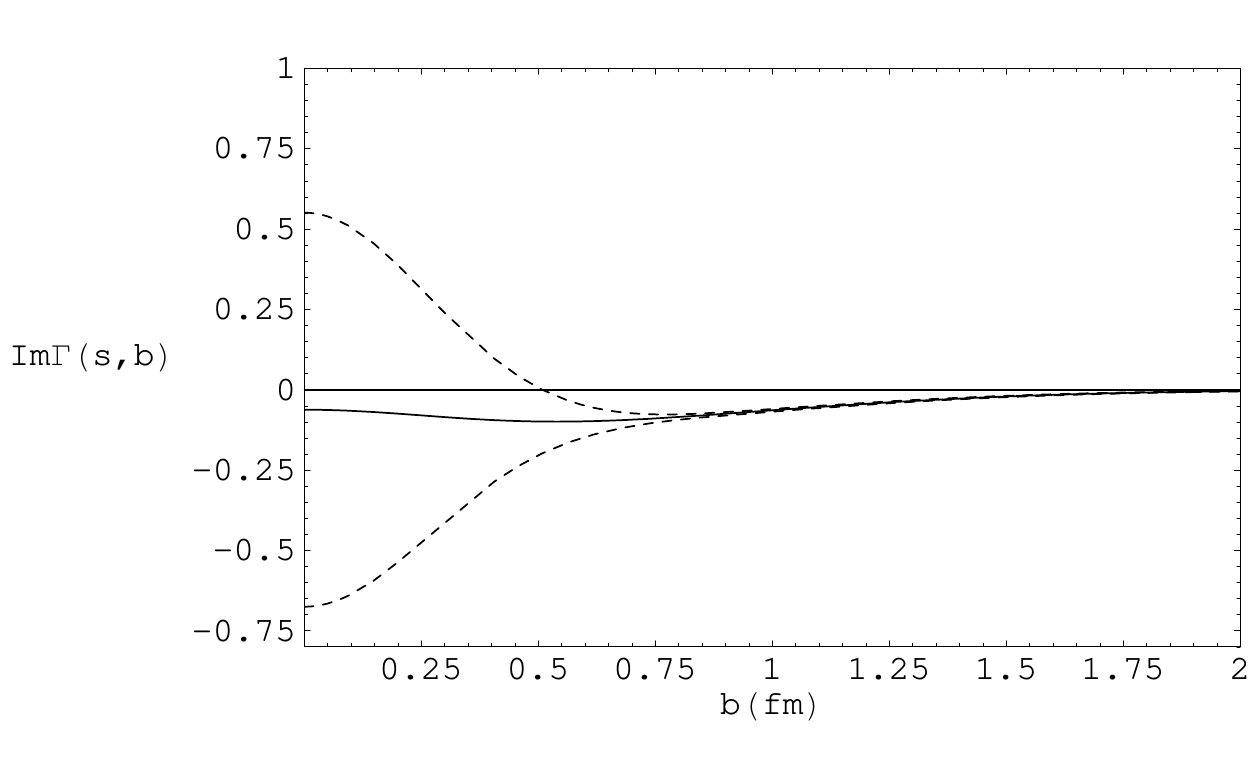}
\vspace*{8pt}
\caption{Real e imaginary parts of the profile function to $\bar{p}p$ at $\sqrt{s}=1800.0$ GeV. Results obtained from the fit parameters to
  ${q^2}_{max}=14.0$ GeV$^2$. Dashed lines represent the error propagation.}
\label{perfil14}
\end{figure*}

The real and imaginary parts of the profile function for $pp$ and $\bar{p}p$ are shown in Figs. \ref{perfil2}, \ref{perfil11} and \ref{perfil14}. The real part of $\mathrm{\Gamma}(s,b)$ is dominant, thus elastic scattering amplitude is almost imaginary at large momentum transfer. Therefore, the elastic scattering is mainly the shadow of the inelastic channels and the momentum transfer dependence reflects the shape of the interacting hadrons \cite{amaldijacob}. 

\subsection{Eikonal function}\label{seikonal}
\begin{figure*}[ht]
\centering
\includegraphics[width=12.0cm, height=10.0cm]{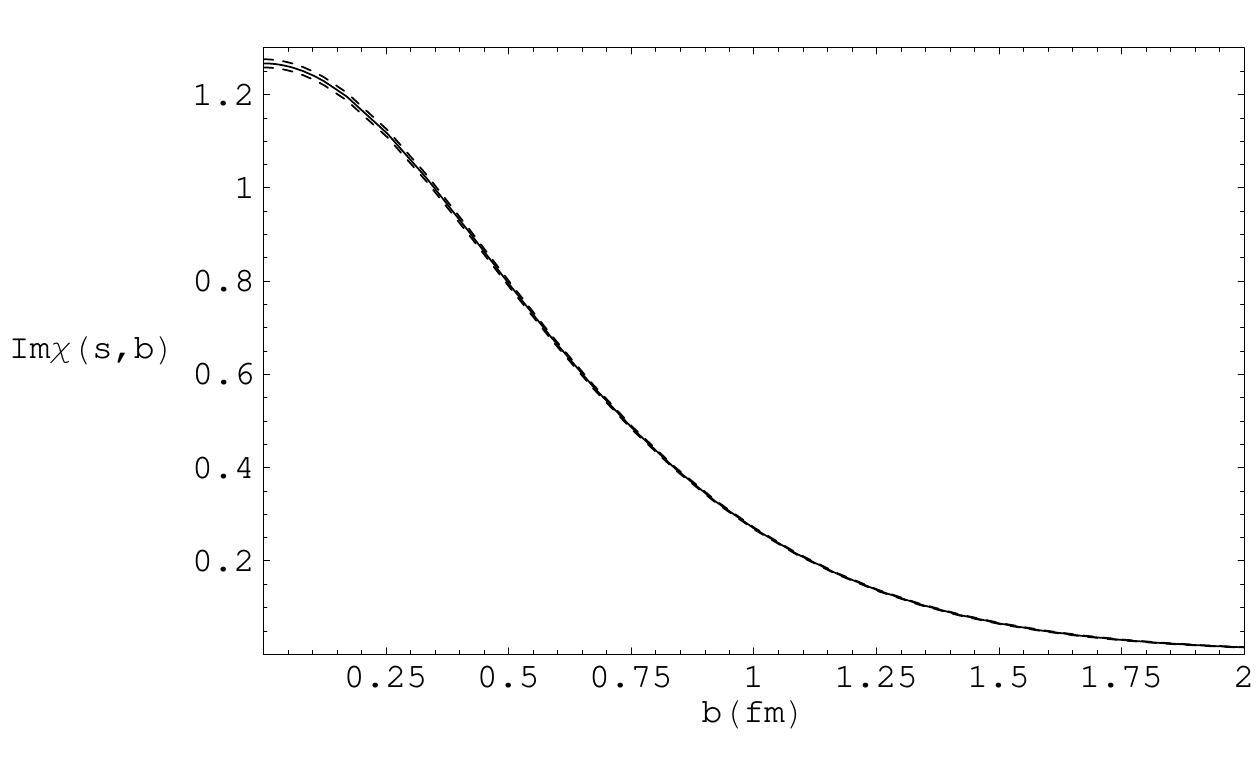}
\includegraphics[width=12.0cm, height=10.0cm]{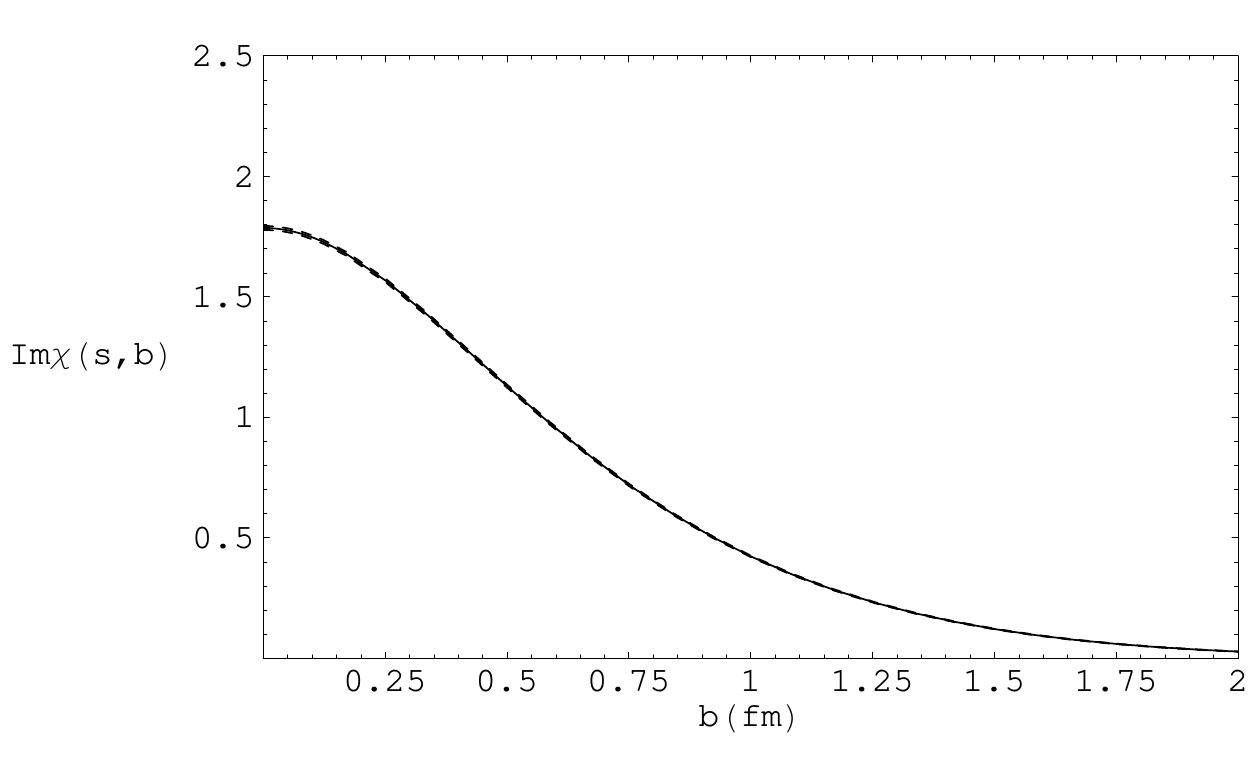}
\vspace*{8pt}
\caption{Imaginary part of the eikonal function to $pp$ at $\sqrt{s}=52.8$ GeV (left) and to $\bar{p}p$ at $\sqrt{s}=546.0$ GeV (right). Results obtained from the fit parameters to ${q^2}_{max}=14.0$ GeV$^2$. Dashed lines represent the error propagation.}
\label{fig:eiconal1}
\end{figure*}

The imaginary part of the eikonal, the so-called opacity, can be viewed in Fig. \ref{fig:eiconal1}. The real part of the eikonal is always positive, which means attenuation and can only come from particle production diagrams \cite{predazzi} not considered in our approach. 

It is interesting to note while not the purpose of this work that the eikonal in the impact parameter space, $\chi(s,b)$ can be connected to eikonal in the momentum transfer space, $\tilde{\chi}(s,q^2)$ by

\begin{eqnarray}
\nonumber \tilde{\chi}(s,q^2)=\int_0^{\infty} bdbJ_0(qb)\chi(s,b),
\end{eqnarray}

\noindent then $\chi(s,q^2)$ allows possible connections with quantum field theory since elementary cross sections are expressed in the momentum transfer space as well as form factors of the nucleons \cite{pas}. In general, the inputs for $\chi(s,q^2)$ are based on analogies with geometry, optics, or microscopic concepts in well established QCD bases. Of course, the crucial test for any model comes from the experimental data on the physical quantities that characterize the elastic scattering. 

\subsection{Inelastic overlap function}\label{overlap}
\begin{figure*}[h]
\centering
\includegraphics[width=12.0cm, height=10.0cm]{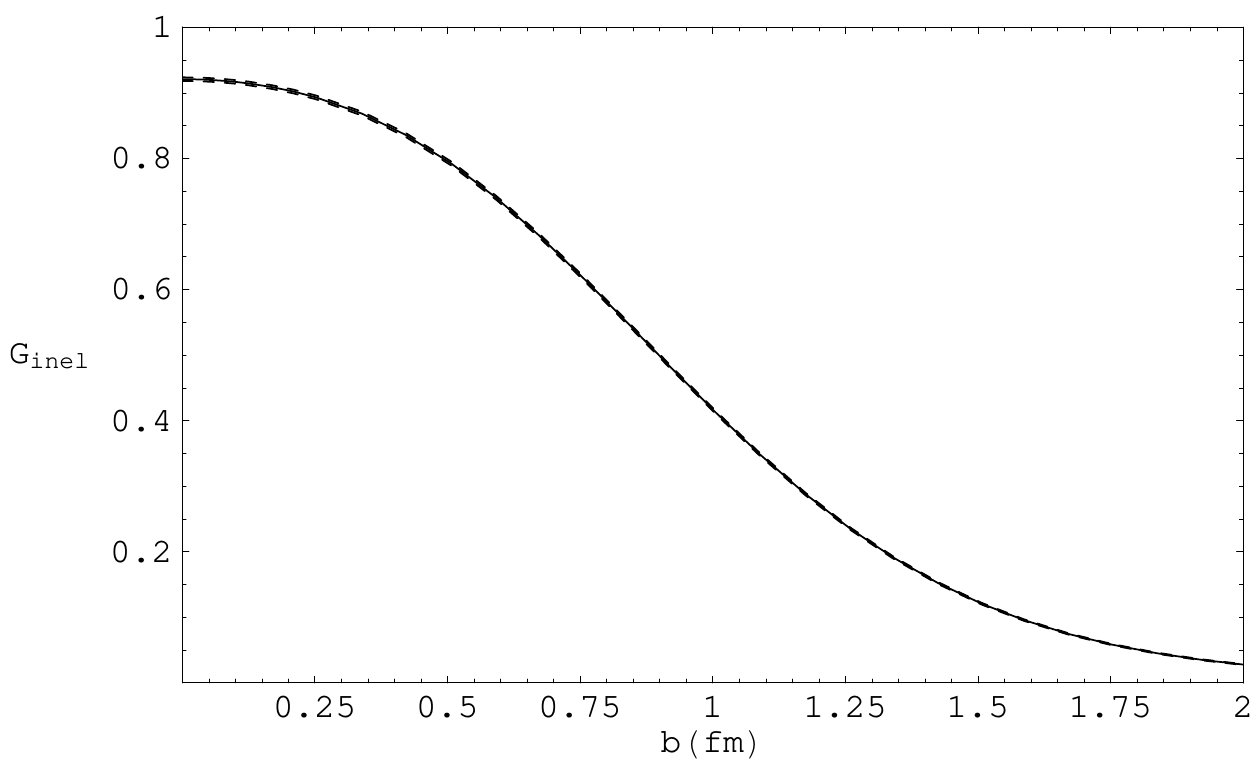}
\includegraphics[width=12.0cm, height=10.0cm]{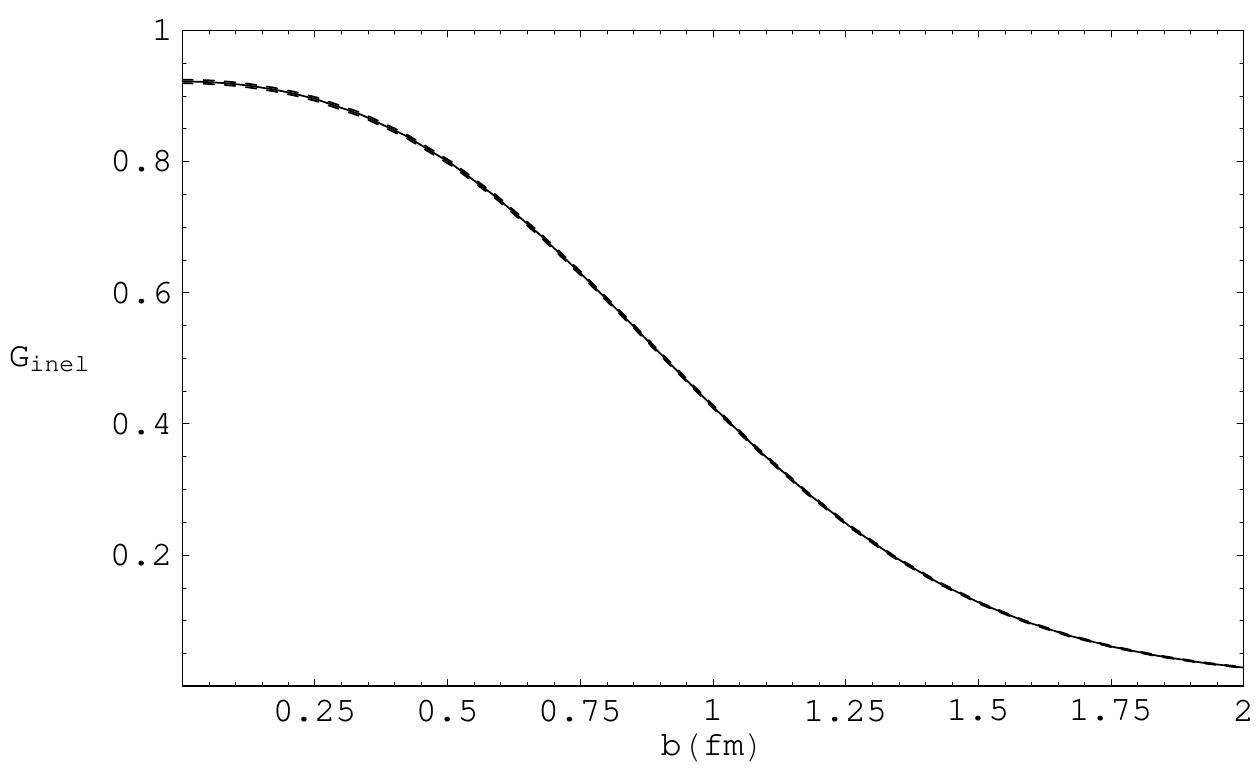}
\vspace*{8pt}
\caption{Inelastic overlap function to $pp$ at $\sqrt{s}=52.8$ GeV (left) and at $\sqrt{s}=62.5$ (right). Results obtained from the fit parameters to ${q^2}_{max}=14.0$ GeV$^2$. Dashed lines represent the error propagation.}
\label{gine1}
\end{figure*}

\begin{figure*}[ht]
\centering
\includegraphics[width=12.0cm, height=10.0cm]{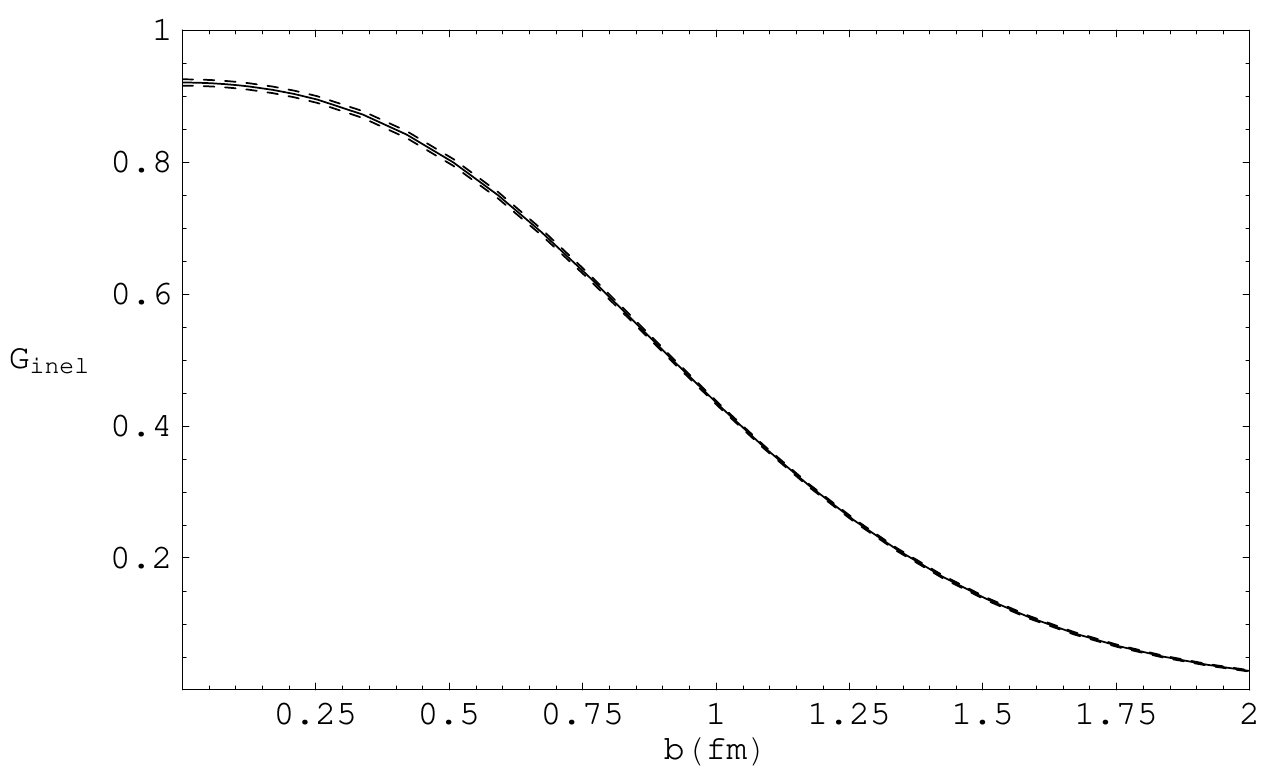}
\includegraphics[width=12.0cm, height=10.0cm]{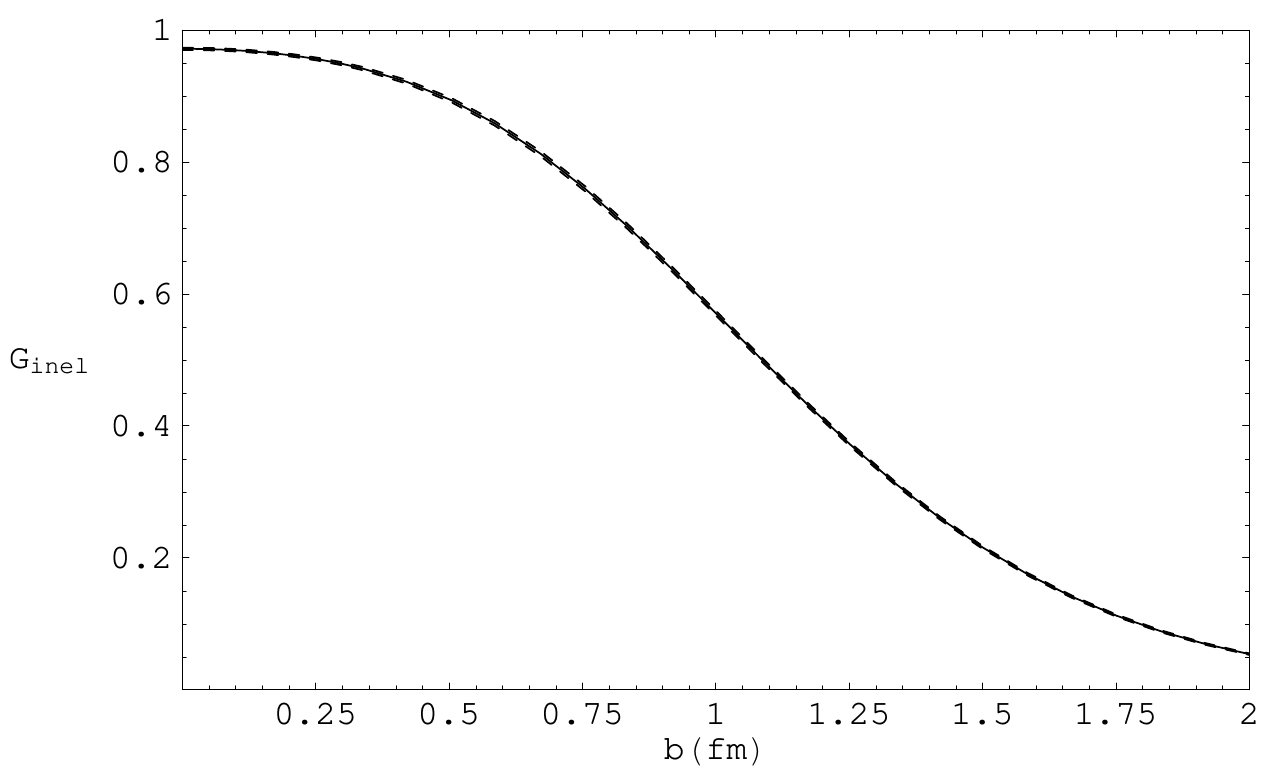}
\vspace*{8pt}
\caption{Inelastic overlap function to $\bar{p}p$ at $\sqrt{s}=53.0$ GeV at ${q^2}_{max}=14.0$ GeV$^2$ (left) and at $\sqrt{s}=546.0$ GeV (right). Dashed lines represent the error propagation.}
\label{gine3}
\end{figure*}

\begin{figure*}[ht]
\centering
\includegraphics[width=12.0cm, height=10.0cm]{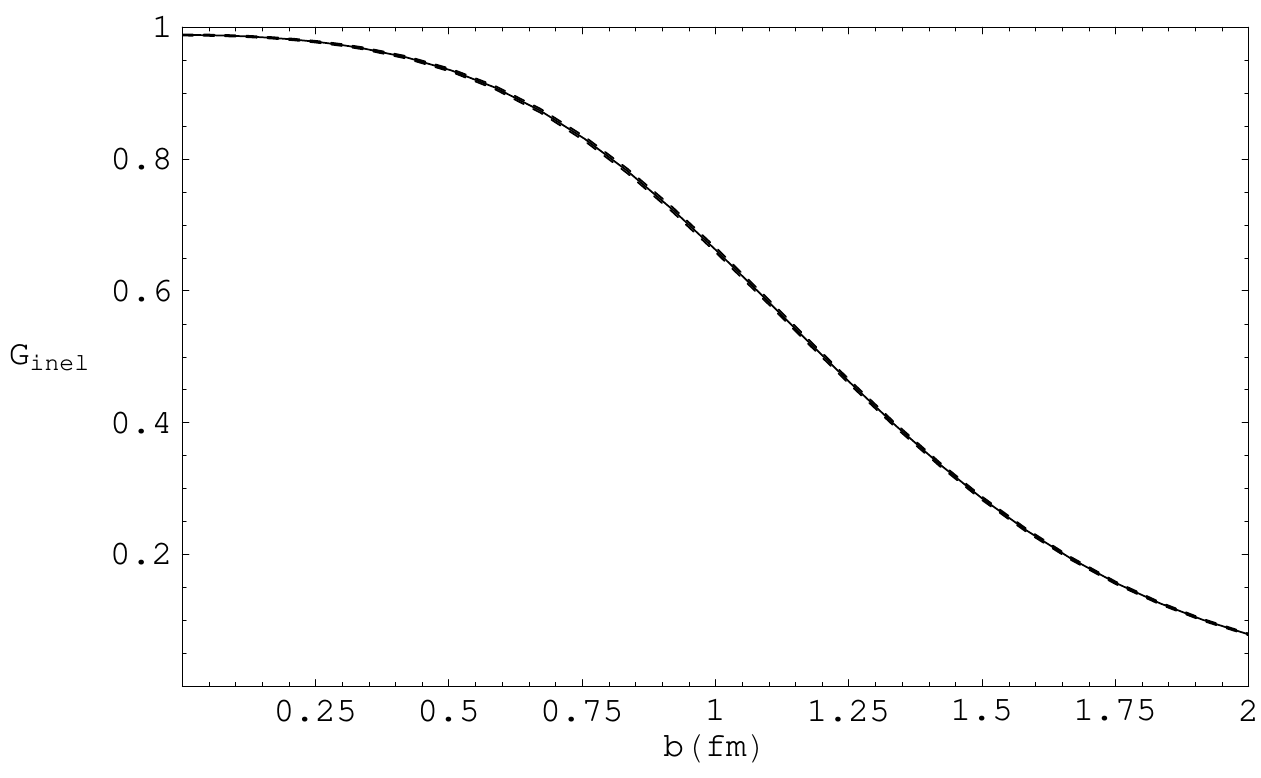}
\vspace*{8pt}
\caption{Inelastic overlap function to $\bar{p}p$ at $\sqrt{s}=1.8$ TeV to ${q^2}_{max}=14.0$ GeV$^2$. Dashed lines represent the error propagation.}
\label{gine5}
\end{figure*}

\begin{figure*}[ht]
\centering
\includegraphics[width=12.0cm, height=10.0cm]{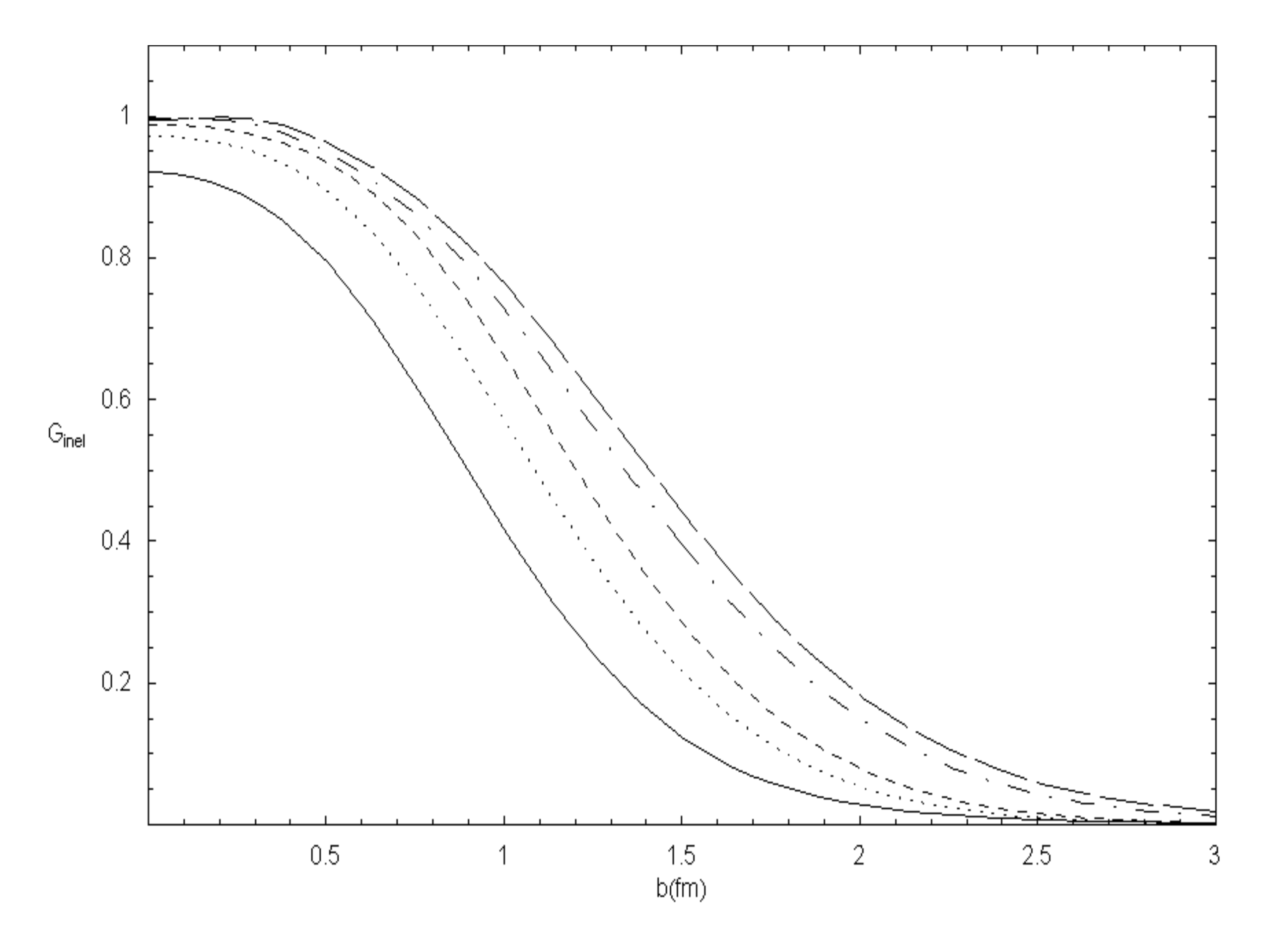}
\vspace*{8pt}
\caption{Inelastic overlap function to: $pp$ at $\sqrt{s}=52.8$ GeV (solid line) (ISR), $\bar{p}p$ at $\sqrt{s}=546.0$ GeV (dotted line) (SPS Collider), $pp$ at $\sqrt{s}=1.8$ TeV (dashed line) (Tevatron), $\bar{p}p$ at $\sqrt{s}=7.0$ TeV (dash-dotted line) (LHC), and $pp$ at $\sqrt{s}=14.0$ TeV (long dashed line) (LHC). The inelastic overlap function grows with increasing energy at each $b$.}
\label{gine6}
\end{figure*}

Amaldi and Schubert pointed out \cite{amaldi} that a precise knowledge of the real part of the scattering amplitude is not required for the impact parameter analysis since the effect of the real part on $G_{inel}(s,b)$ is negligible. The inelastic overlap function can be viewed in Figs. \ref{gine1}, \ref{gine3} and \ref{gine5}. In Fig. \ref{gine6} are displayed the inelastic overlap functions to $pp$ at $\sqrt{s}=52.8$ GeV, $\sqrt{s}=7.0$ TeV, $\sqrt{s}=14.0$ TeV, and to $\bar{p}p$ at $\sqrt{s}=546.0$ GeV, and at $\sqrt{s}=1.8$ TeV. Notice that the overlap function grows with the energy increase, but it is a non-increasing function of $b$ for all smaller values of $b$. The Dubna Dynamical Model \cite{dubna} of hadron-hadron scattering at high energy is based on the structure of a hadron as a compound system with a central region which the valence quarks are concentrated and log-distance region filled with a color-singlet quark-gluon field. As a result, the hadron amplitude can be represented as a sum of a central and a peripheral part and predicts the saturation of the overlapping function starting at $\sqrt{s}\sim 1.8$ TeV, as obtained here.

Contrary to Kundrat and Lokajicek work \cite{kundrat,adachi}, the inelastic overlap showed here will be non-negative for any value of $b$. That occurs because the scattering amplitudes (\ref{espalhamentopp}) and (\ref{espalhamentopbp}) are non-oscillating functions around zero at high values of $b$. Furthermore, Kundrat and Lokajicek define their inelastic overlap function, $G_{inel}^{KL}(s,b)$, using unitary condition in terms of the impact parameter amplitude 

\begin{eqnarray}
\nonumber h_{el}(s,b)=\frac{\eta(s,b)\exp(i\mathrm{Re}\chi(s,b)) -1}{2i},
\end{eqnarray}

\noindent where $h_{el}(s,b) = \exp(- \mathrm{Im}\chi(s,b))$. On the other hand, here $G_{inel}(s,b)$ is defined using unitary condition in terms of the profile function

\begin{eqnarray}
\nonumber \Gamma(s,b) = 1 - \eta(s,b)\exp(i\mathrm{Re}\chi(s,b))
\end{eqnarray}

\noindent resulting $G_{inel}(s,b)^{KL}=G_{inel}(s,b)/4$. The saturation of the Froissart-Martin bound may not be reached at LHC energies using this approach.

\section{Predictions}\label{predictions}
\begin{figure*}[b]
\centering
\includegraphics[width=12.0cm, height=10.0cm]{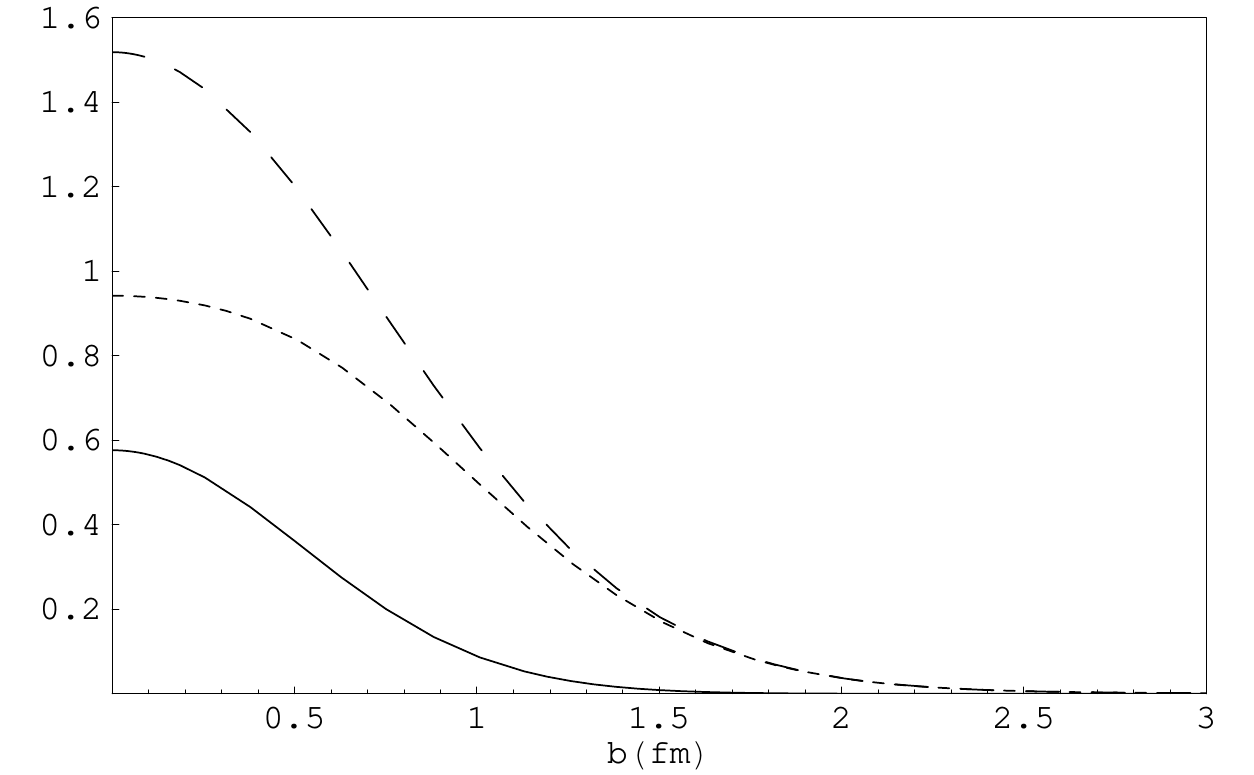}
\vspace*{8pt}
\caption{Comparison between $G_{inel}(s,b)$ (dotted line), 2$\mathrm{Re}\mathrm{\Gamma}(s,b)$ (dashed line) and $|\mathrm{\Gamma}(s,b)|^2$ (solid line) to $pp$ at $\sqrt{s}=$200.0 GeV. Results obtained from the fit parameters to ${q^2}_{max}=14.0$ GeV$^2$.}
\label{gine10}
\end{figure*}

\begin{figure*}[t]
\centering
\includegraphics[width=12.0cm, height=10.0cm]{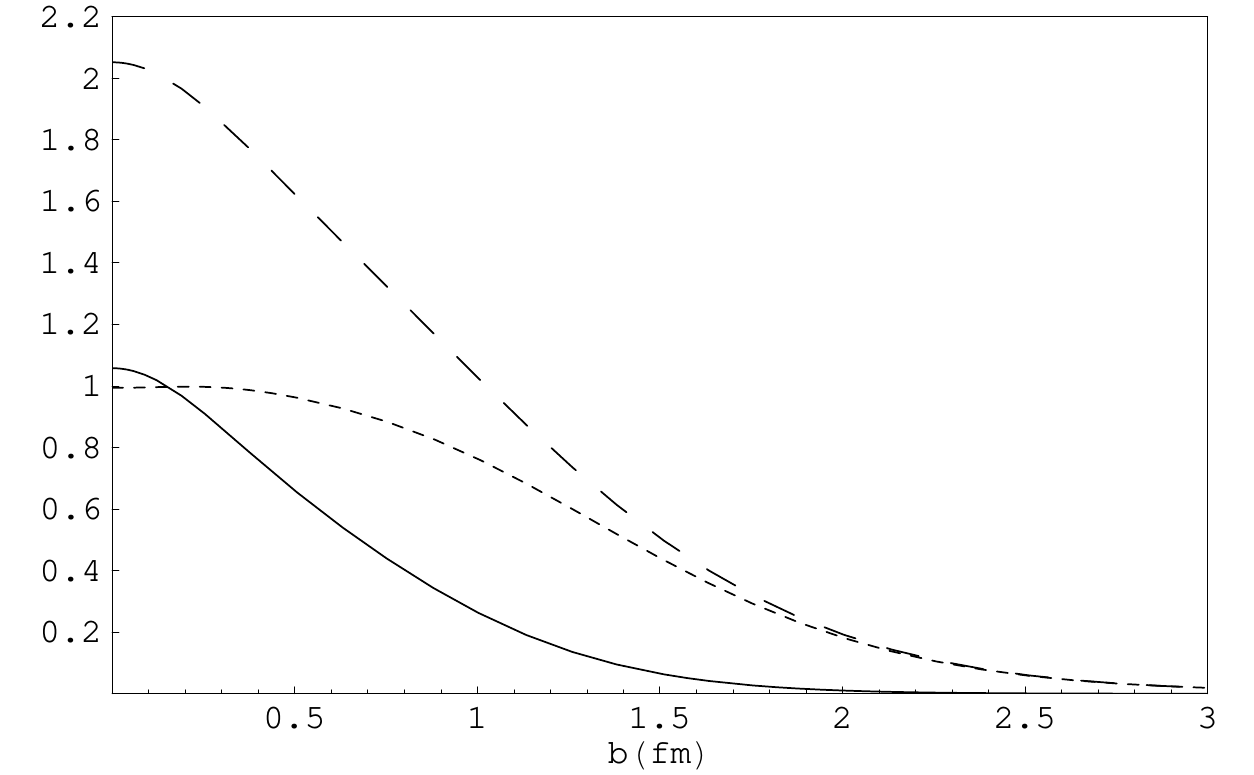}
\vspace*{8pt}
\caption{Comparison between $G_{inel}(s,b)$ (dotted line), 2$\mathrm{Re}\mathrm{\Gamma}(s,b)$ (dashed line) and $|\mathrm{\Gamma}(s,b)|^2$ (solid line) to $pp$ at $\sqrt{s}=$14.0 TeV. Results obtained from the fit parameters to ${q^2}_{max}=14.0$ GeV$^2$.}
\label{gine11}
\end{figure*}

Figs. \ref{gine10} and \ref{gine11} show predictions for $|\mathrm{\Gamma}(s,b)|^2$, $2\mathrm{Re}\mathrm{\Gamma}(s,b)$, and $G_{inel}(s,b)$ at energies not available in the fit procedure described in \cite{sdc}. 

At $\sqrt{s}=14$ TeV, the inelastic overlap function almost reaches the saturation, i.e., $G_{inel}(s,b)\approx 1$. From (\ref{eq10}), we may conclude that $\mathrm{Im}\chi(s,b)\rightarrow \infty$ and so the total cross section indicates the saturation of the Froissart-Martin bound. This saturation at $\sqrt{s}=14.0$ TeV may imply corrections in the formalism used to describe this kind of scattering process. Therefore, some novel physical mechanism must be used to establish the correct growth of the elastic total cross section, if LHC run confirms this prediction.  

On the other hand, $|\mathrm{\Gamma}(s,b)|^2=[\mathrm{Re}\mathrm{\Gamma}(s,b)]^2+[\mathrm{Im}\mathrm{\Gamma}(s,b)]^2$ and from Fig. \ref{gine11} we see that $\mathrm{Re}\mathrm{\Gamma}(s,b)\approx 1$ and $\mathrm{Im}\mathrm{\Gamma}(s,b)$ is very small at $b=0$. Therefore, the profile function is real dominant, i.e., the elastic scattering amplitude in the momentum transfer space will be imaginary dominant. Consequently, we expect a negligible real part of the scattering amplitude at $\sqrt{s}=14$ TeV for $q^2$ small values.

\section{Final Remarks}\label{final}
Empirical parameterization of the elastic scattering amplitude and fits of the available data have widely been used as a source of model-independent analysis and constitute important tactics that can contribute with the establishment of novel theoretical calculational schemes. The approach here applied is used to determine several quantities of interest, such as the profile, the eikonal, the inelastic overlap functions. With some additional hypotheses, even information about the form factors (momentum transfer space) can be extracted. 

Results in the impact parameter representation were obtained using the Fourier-Bessel transform of the elastic scattering amplitude in the momentum transfer space. In the impact parameter representation, we obtain the profile, eikonal, and the inelastic overlap functions in a confidence interval due to standard error propagation in the fit parameters. All results in the parameter space representation show that the maximum of their distributions lies at positive values of the impact parameter $b$. In the parameterization proposed in \cite{sdc} an important assumption is that real part of the scattering amplitude may be neglected in comparison to the imaginary part at sufficient high values of transferred momentum. This assumption implies in a peripheral character to the $pp$ and $\bar{p}p$ elastic processes, i.e., it occurs at small $b$ (central character).

In the almost model-independent analysis performed here, $G_{inel}(s,b)$ is practically saturated at $\sqrt{s}=14$ TeV. Therefore, future experimental data obtained at LHC could be compared with this result allowing a confidence test of the parameterization. Furthermore, this premature saturation may show some novel physical mechanism in this energy region.

As a remark, we note that the impact parameter description does not claim to be able to fit the data quantitatively. However, it can provide
evidence of fundamental features of high-energy scattering contributing for further developments of more elaborate diffractive models.

\section*{Acknowledgment}
The author is thankful to UFSCar for financial support.

\appendix

\section{Uniqueness of the Profile Function}
Consider the known elastic scattering amplitude $F(s,q^2)$ written as a Fourier-Bessel transform in the range $(0\leq q^2 < q_{max}^2)$
\begin{eqnarray}
\nonumber F(s,q^2)=i{\int_0}^{\infty}bdbJ_0(qb)\mathrm{\Gamma}(s,b).
\end{eqnarray}

The profile function is now known. On the other hand, consider the unknown amplitude $\tilde{F}(s,q^2)$ with respective unknown profile function
\begin{eqnarray}
\nonumber \tilde{\mathrm{\Gamma}}(s,b)=i\int_{q_{max}^2}^{\infty}q'dq'J_0(bq')\tilde{F}(s,{q'}^2).
\end{eqnarray}

The combined profile function is $\mathrm{\Gamma}(s,b)+\tilde{\mathrm{\Gamma}}(s,b)$ and for $(0\leq q^2< q_{max}^2)$ we obtain the known amplitude
\begin{eqnarray}
\nonumber F(s,q^2)=i{\int_0}^{\infty}bdbJ_0(qb)[\mathrm{\Gamma}(s,b)+\tilde{\mathrm{\Gamma}}(s,b)].
\end{eqnarray}

On the other hand, taking $(q_{max}^2<q^2<\infty)$ we obtain
\begin{eqnarray}
\nonumber \tilde{F}(s,q^2)=i{\int_0}^{\infty}bdbJ_0(qb)[\mathrm{\Gamma}(s,b)+\tilde{\mathrm{\Gamma}}(s,b)],
\end{eqnarray}

\noindent the unknown amplitude. Of course, we may construct an arbitrary number of scattering amplitudes (or profile functions) each one well established in its range. Therefore, for the sake of simplicity we assume $F(s,q^2)=\tilde{F}(s,q^2)$ and hence, the profile function obtained in this work is unique.


\end{document}